\documentclass[aps,preprint,showkeys,nofootinbib,prb]{revtex4-1}%
\usepackage{amssymb}
\usepackage{amsfonts}
\usepackage{amsmath}
\usepackage{graphicx}
\usepackage{natbib}%
\providecommand{\U}[1]{\protect\rule{.1in}{.1in}}
\setcitestyle{numbers}
\setcitestyle{square}

\begin{document}
\title{Entanglement and confinement: A new pairing mechanism in high-$T_{C}$ cuprates}
\author{Felix A. Buot$^{1,2}$, Roland E. S. Otadoy$^{2}$, and Unofre Pili$^{2}$}
\affiliation{$^{1}$C\&LB Research Institute, Carmen 6005, Cebu, Philippines, }
\affiliation{$^{2}$LCFMNN, TCSE Group, Department of Physics, University of San Carlos,
Talamban 6000, Cebu, Philippines}

\begin{abstract}
We demonstrate that entanglement and confinement hole pairing (ECHP) is a
precise physics of the entanglement framework of the RVB theory of
high-$T_{C}$ cuprates. Our novel strong ECHP mechanism explains the entire
phase diagram of both electron and hole-doped cuprates, notably the linearly
decreasing $T^{\ast}$ at the pseudogap but with "$T^{\ast}$-singularity" at
the peak of the superconducting (SC) dome, the $T_{C}=T^{\ast}$ at the optimum
doping and the rest of the overdoped regions of the SC dome, the duality of
the spin gap and strange metal phase, the presence of the parallel
superconducting stripes in the CuO plane (spin-polarized and spin-unpolarized
channels), and the linear-$T$ resistivity of the strange metal phase above the
overdoped regions of the SC dome. This also explains the experimental spin
textures of the cuprates. We refer to our new ECHP model as a resonating
entanglement and confinement hole pair (RECHP) theory. Based on RECHP theory,
we were able to provide a conceptual and comprehensive semiquantitative
explanation of the entire phase diagram, thus providing the sought-after
pairing mechanism responsible for the entire phase diagram of high-$T_{C}$ cuprates.

\end{abstract}
\maketitle

\section{Introduction}

The purpose of this study is to analyze the typical phase diagram of
high-$T_{C}$ cuprates based on our new ECHP mechanism. We argued that the
predictions of our pairing model completely agree with the experimental phase
diagrams of cuprates. We resolved how this new ECHP mechanism is responsible
for the entire phase diagram of both electron and hole doped cuprates. For the
first time, we unraveled the singularity of the $T^{\ast}$-curve at the SC
dome peak, compatible with the experimental phase diagram of cuprates. This
singularity cannot occur in most theories which yield analytic $T^{\ast}%
$-curve. We pointed out that the inherent entanglement framework of the
original RVB theory of high-$T_{C}$ cuprates must be reformulated to introduce
the concept of confinement of doped holes for precise physics.

\begin{figure}
[ht!]\centering\includegraphics[width=5.785900in]{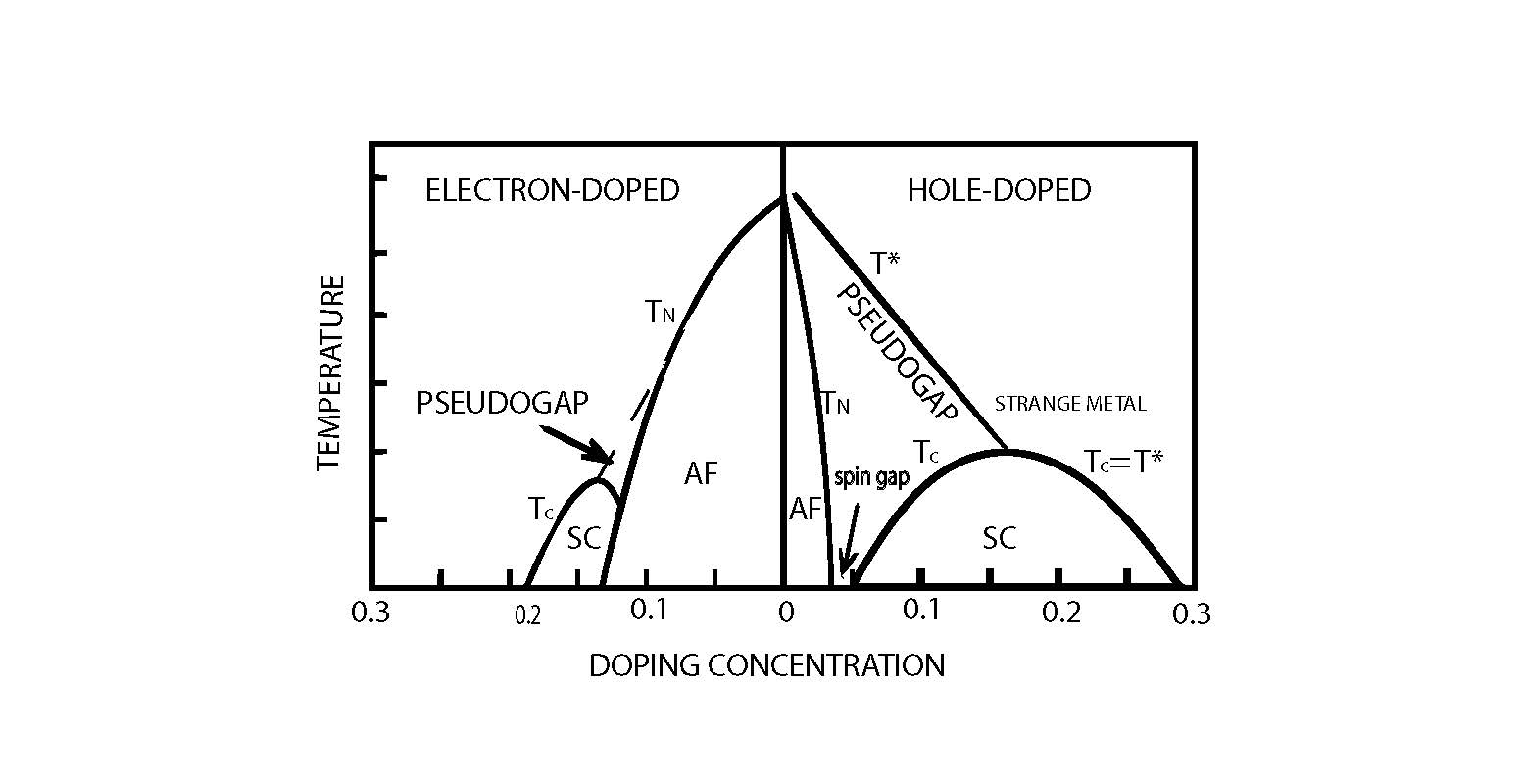}
\caption{Generic phase diagram for the high $T_{C}$ superconductors. Our theory produces the spin gap, the singularity of $T^{\ast}$
at the SC dome peak for both the electron and hole doped cuprates and is consistent with $T_{C}=T^{\ast}$ at the SC dome
peak and down through the overdoped region of the SC dome. [Figure redrawn and edited from Ref. \cite{edeger}]}
\label{phaseD}
\end{figure}

Succinctly speaking, the phase diagram, Fig. \ref{phaseD}, is mainly
characterized by two order parameters, namely, (1) the preformed pair order
(PO), that is, condensation to lower energies at $T^{\ast}$ of the disordered
"nematic" preformed entangled pairs and (2) the "smectic" configurational
order (CO), that is, a global symmetry-breaking (SB) transition from "nematic"
to "smectic" ordering at $T_{C}$ of the SC dome, eventually adhering to
$T^{\ast}=T_{C}$ in the overdoped region of the SC dome. The graph for
$T^{\ast}$ is deduced analytically, while the SC dome is obtained graphically.
The evolution process from $T^{\ast}$(PO) to $T_{C}$ (CO) is described by the
transition rate $R=\frac{\Delta\left(  CE\right)  }{\Delta T}$, where CE is
the configuration entropy (CE). We note that at CO, CE=0, since CO signifies a
unique arrangement of long-distance entangled pairs. The spin gap is
characterized by the failure to attain CO, after PO at $T^{\ast}$, upon
continued cooling down to a limiting lowest temperature because the rate $R$
is insufficient to attain CO. On the other hand, at the peak of the SC dome,
$R$ goes to infinity. The strange metal phase is characterized by the
destruction of PO but with surviving CO at $T>T_{C}$. Hence, spin gap and
strange metal phases have complementary or dual properties. First, we place
our research in a much wider perspective in what follows.

\subsection{The search for a unifying theory of cuprates}

The Bardeen--Cooper--Schrieffer (BCS) theory of superconductivity developed in
the late 1950 is an extremely successful paradigm for understanding
conventional superconductors. The consensus among theoretical physicists
\cite{keimer, anderson4} seems to be that phonon mediated pairing of electrons
yields $T_{C}$ far below the maximum $T_{C}$ of copper oxides. Magnetic
excitations mechanisms have also been vigorously explored owing to the success
of BCS. However, excitations or boson-mediated pairing to produce composite
bosonic-charge quasiparticles that condense into the superfluid state have
proven to be inadequate for explaining, predicting, and elucidating the entire
phase diagram of high-$T_{C}$ cuprates.

The SC transition temperatures of copper oxides were discovered in 1986
\cite{bednorz}. The maximum $T_{C}$ significantly exceeds that of any
previously known superconductor. In fact, for HgBaCaCuO under pressure
\cite{keimer}, the highest $T_{C}\simeq165K$ obtained at that time. The
\textit{stripy} pattern of \textit{unidirectional} $CuO$ planar conduction in
superconducting states, as observed in scanning tunneling spectroscopy (STS)
\cite{nanjing, ref1}, appears mysterious. \ Moreover, the origin of the
complex spin texture of some copper oxides obtained by more recent
spin-resolved ARPES remains heuristic or empirical \cite{gotlieb,iwasawa,lou}.
The holy grail lies in the search for a strong pairing mechanism responsible
for the high-$T_{C}$ of copper oxides. It is believed that this new strong
pairing mechanism is responsible for the entire phase diagram of high-$T_{C}$
cuprates \cite{singh2}. In this study, we propose the sought-after pairing
mechanism responsible for the entire phase diagram, for both hole- and
electron-doped high-$T_{C}$ cuprates.

\subsection{Crystal structure and dynamical degrees of freedom of cuprates}

\begin{figure}
[ht!]%
\centering\includegraphics[width=6.590700in]{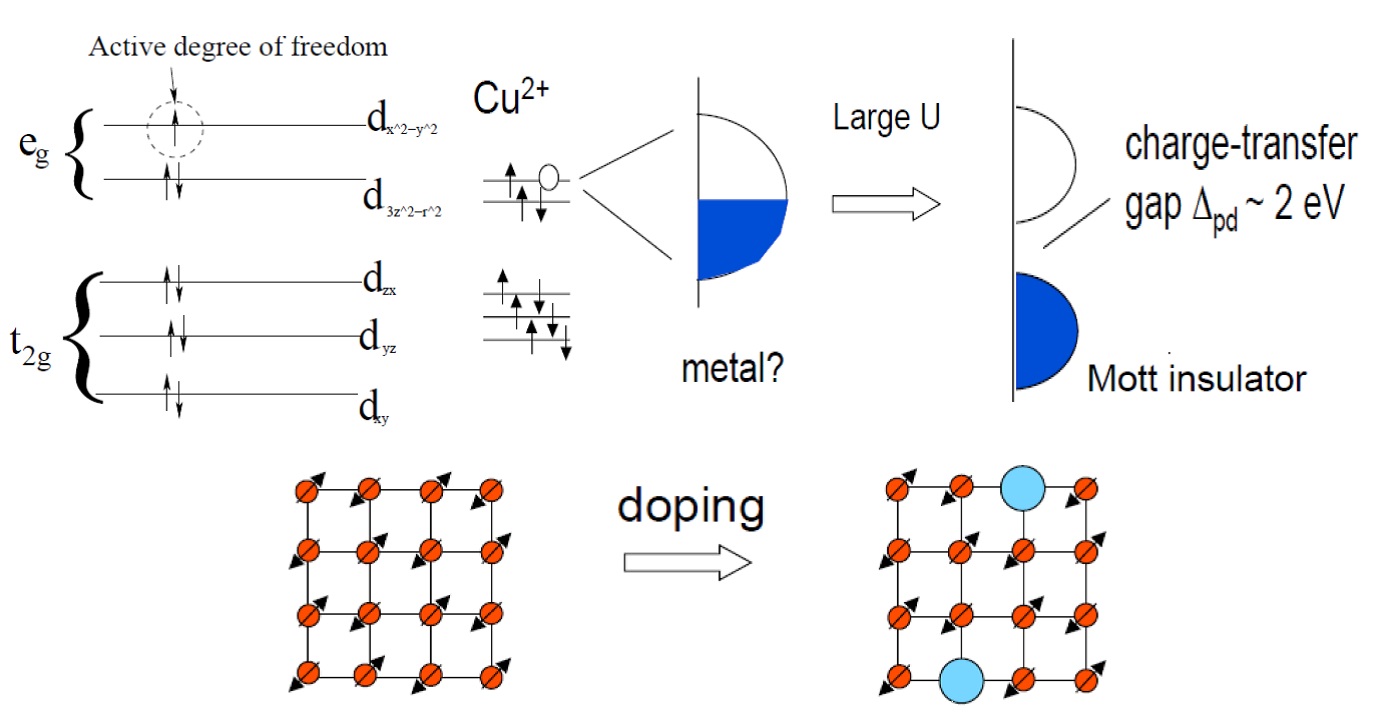}\caption{ Antiferromagnetic cuprates are Mott insulators in the undoped state due to strong electron-electron correlation
and Coulomb blockade between Cu sites. In the cuprates, carrier doping of the Mott insulating parent state is
necessary to realize superconductivity. We assume that the antiferromagnetic-chain link between entangled pair
is still in the Mott insulating state, that is the strong electron-electron correlation is still accounted for.  Dopant holes
(big dots) can reside in the Cu sites for hole-doped cuprates. The figure shows the relevant degrees of freedom
responsible for both antiferromagnetism and superconductivity.
[Figure taken from different sources, namely, \cite{singh2} and \cite{norman}]}
\label{mott}
\end{figure}

Cuprates are layered materials of superconducting copper oxide planes,
separated by layers of dopants such as lanthanum, barium, strontium, and
doping electrons or holes into the copper-oxide planes. The crystal structure
of cuprates is lamellar, with two CuO2 layers sandwiched between spacer layers
\cite{anderson2}. The active element where conduction occurs is the $CuO_{2}$
layer. Typical cuprates have an octahedral cage that is elongated along the
c-axis owing to Jahn-Teller distortion, for example, LSCO \cite{anderson2}.
The degrees of freedom in the electronic structure are depicted in Fig.
\ref{mott}.

The $d^{9}$ orbitals of the undoped copper atom are split into two sub-groups,
$e_{g}$ and $t_{2g}$, owing to crystal field effects \cite{singh2}. Subgroup
$t_{2g}$ forms the lower energy set and $e_{g}$ forms the higher energy set.
One electron entered the $d_{x^{2}-y^{2}}$ orbital. Thus, the $d_{3z^{2}%
-r^{2}}$ orbital is in the lower-energy state and $d_{x^{2}-y^{2}}$ orbital is
in the higher-energy state. Hybridization does not radically modify this
picture \cite{anderson2}. Thus, the $d_{x^{2}-y^{2}}$ orbital is responsible
for both magnetism (in the undoped state) and superconductivity (in the doped
state) \cite{anderson2, anderson3}. We concurred with Anderson's view that
\textit{upon doping, electrons are removed from the Cu sites} \cite{anderson2}%
. Others believe that hole doping removes electrons from oxygen $p$ orbitals,
in two sub-systems: itinerant and localized \cite{barzy}. This view remains
debatable \cite{singh}. Recent studies have shown that during electron doping
holes are generated at the oxygen sites.

\subsection{Nonperturbative RVB and BCS-like theories}

The physical picture of the original idea of the RVB state \cite{anderson2} is
that strong correlations of the unpaired electrons in $d_{x^{2}-y^{2}}$ copper
orbitals are localized in the undoped state yielding a Mott insulator state.
The spins of unpaired electrons form singlet pairs with nearby neighbors
\cite{singh2,pauling}. In $d$-wave pairing state, electrons simply avoid being
very close thus Coulomb repulsion is reduced \cite{anderson2,anderson3}.The
physical picture that emerges from RVB and related theories is that the
\textit{superexchange interaction} $J$ \textit{is the cause of pairing
\cite{singh2}}. A pseudogap is a phase disordered superconductive dimers for
hole-doped cuprates. The continuous $T^{\ast}$across the optimal doping or
peak of the SC dome is not supported by experiments and is a major weakness of
the RVB and other RVB-based theories \cite{singh2, marino}. RVB plain-vanilla
is a partially successful theory \cite{singh2,edeger}. One crucial result of
our present study is the presence of a mathematical singularity of $T^{\ast}%
$curve, producing a "kink" at the peak of the SC dome compatible with
experimental results.

There are magnetic fluctuation-exchange theories but has limited success in
explaining the phase diagram of cuprates \cite{moller,magnon}.
Paramagnon-pairing theories are not based on superexchange $J$
\cite{monien,sushkov} unlike RVB. All other magnetic excitation-based theories
have problems based on energetics considerations \cite{singh2}. Slave-boson or
slave-fermion techniques for strongly-correlated systems have been applied but
have not been fully successful in high-$T_{C}$ cuprates \cite{chui,chan}.
Topological theories involving the gauge and Chern-Simons theories have also
been applied. Spin-polaron and bipolaron theories are some of the active
fields of research \cite{wood,yanga,Alexandrov} but they have not yet
completely explained the entire phase diagram, for both electron- and
hole-doped cuprates. In summary, there are no well-accepted unifying physical
concept and theory of high-$T_{C}$ cuprates for both electron and hole doped.

\section{A novel ECHP mechanism: confinement of holes}

Here, we propose\ a doping-dependent long-distance pairing of doped holes
based on quantum entanglement and confinement. In this new pairing mechanism,
determined by superexchange bonding, $J$, the coupling strength increases
linearly with distance $L$ of the antiferromagnetic-chain link between the
hole pairs. This is precisely our definition of confinement, which is
controlled by doping levels. This is \textit{akin} to the confinement of
quarks by gluons in quantum chromodynamics. Our theory strongly suggests a
drastic conceptual and physical reformulation of Anderson's RVB theory,
although both theories are fundamentally based on quantum entanglements.
Strong pairing entanglement comes, in the quantum information sense, as
a\ summation of the entanglement entropy of formation (EEF) calculated for
each disentangled unit \cite{perspective,buotbook2}. This is derived in what follows.

\subsection{A direct measure of confinement}

As one would expect, the degree of entanglement will convey the strength of
the coupling between a pair. The two properties of any entangled qubit,
namely, \textit{concurrence} and its \textit{emergent} qubit
\cite{perspective,buotbook2}, led Wooters \cite{wooters1, wooters2} to
introduce a direct measure of the degree of entanglement, or confinement as
applied in our present study. This is incorporated in the two formulas,%
\begin{equation}
E\left(  C\right)  =H\left(  \frac{1}{2}+\frac{1}{2}\sqrt{1-C^{2}}\right)  ,
\label{concur}%
\end{equation}
where $C$ is the \textit{concurrence} \cite{perspective} and $H$ is the
Shannon entropy function,%
\begin{equation}
H\left(  x\right)  =-x\ln x-\left(  1-x\right)  \ln\left(  1-x\right)  .
\label{qubit}%
\end{equation}
Concurrence is defined by the invariance of entangled states, such as the Bell
entangled basis states, when all spins are flipped. The global phase factor
does not change the quantum state. For maximally entangled qubits, $C=1$,
which is defined by the invariance of the state by flipping all spins.
Equation (\ref{concur}) states that if there is complete concurrence, that is,
$C=1$, Eq. (\ref{qubit}) states that the system behaves as an emergent qubit
or two-state system. Thus, for \textit{maximally} entangled multi-partite
qubits, we have,%
\begin{align}
E\left(  C\right)   &  =H\left(  \frac{1}{2}\right)  ,\label{neq72}\\
H\left(  \frac{1}{2}\right)   &  =\frac{1}{2}\ln2+\frac{1}{2}\ln2,\nonumber\\
&  =1. \label{neq73}%
\end{align}
affirmed that entangled qubits, either pairs or multi-qubits, behave as a
two-state system or as an \textit{emergent qubit}, yielding an entanglement
entropy equal to one. For further details, please refer to \cite{perspective,
buotbook2}.

\subsection{Strong coupling of RECHP than RVB}

The pair-coupling interaction between hole pair is antiferromagnetic both
vertically and horizontally. Other directions besides the horizontal and
vertical diections do not readily support antiferromagnetic-chain links in
$CuO_{2}$ 2D plane. The coupling of holes by superexchange $J$ at both ends of
an antiferromagnetic chain is our exact definiton of entangled pair since what
one does at one end will immediately be communicated to the other end of the
chain. In the 2D $CuO_{2}$ plane, only vertical and horizontal
antiferromagnetic chains are present, as shown in Fig. \ref{RECPdiag}. Holes
separated by distances across vertical or horizontal antiferromagnetic chains
are very restrictive and cannot be coupled by antiferromagnetic-chain links of
arbitrary length, hence they are not considered according to our definition of
entanglement. Indeed, the singlets in RVB theory are also restricted to
vertical or horizontal directed singlets, as shown in Fig. \ref{RVBdiag}.

\begin{figure}
[ht!]\centering\includegraphics[width=3.397800in]{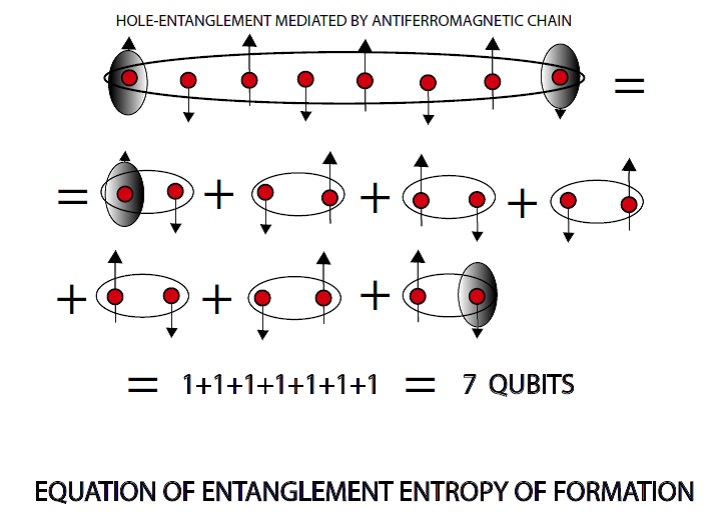}
\caption{ In the figure, for example, the EEF is calculated to amount to seven qubits. The separation process is done
one qubit at a time and is based on the concept of emergent qubit for maximally entangled systems. This
disentanglement process is done seven times to obtain the total EEF. It follows that the entanglement
pairing gap, $\Delta^{\ast}=7\times J$ where $J$ is the superexchange bonding.}
\label{figchain_reduce}
\end{figure}

Figure \ref{figchain_reduce} shows that the EEF of a longer chain is larger
than that of the two nearest qubits. Indeed, the EEF of a nearest-neighbor
pair of entangled qubits is only one qubit. As expected, the entanglement
entropy of formation is directly related to the superexchange bonding, $J$.

The understanding of Fig. \ref{figchain_reduce} revolves around the concept of
an \textit{emergent} qubit \cite{perspective}. The basic principle is that a
maximally entangled system is equivalent to a single qubit. All maximally
entangled qubits exhibited a concurrence $C=1$. This implies that the entire
maximally entangled system often acts as a single qubit.

The physical operation of a series of disentangling steps to calculate the
total EEF is based on the idea of successively breaking out one qubit at a
time. Note that this qubit is entangled with an \textit{emergent} qubit
defined by the segment of the remaining antiferromagnetic chain. We observed
that the total EEF of the series of unentangling processes, one qubit at a
time, amounts to summing up the bonding $J$ between neighboring spins; hence,
the reason for the multiplication with the exchange $J$. Symbolically, we may
write to emphasize the huge difference between RECHP and RVB in terms of
pairing potential as,%
\begin{align}
\Delta_{RECHP}  &  =%
{\displaystyle\sum}
\Delta_{RVB}\label{compare1}\\
EEF\times J  &  =%
{\displaystyle\sum_{L_{B}}}
J \label{compare2}%
\end{align}
where $L_{B}$ is the bond length or is the distance between a pair of
nearest-neighbor Cu sites in a antiferromagnetic chain of the CuO plane.

\subsection{RVB versus RECHP: commonalities and differences}

\begin{figure}
[ht!]\centering\includegraphics[width=4.937200in]{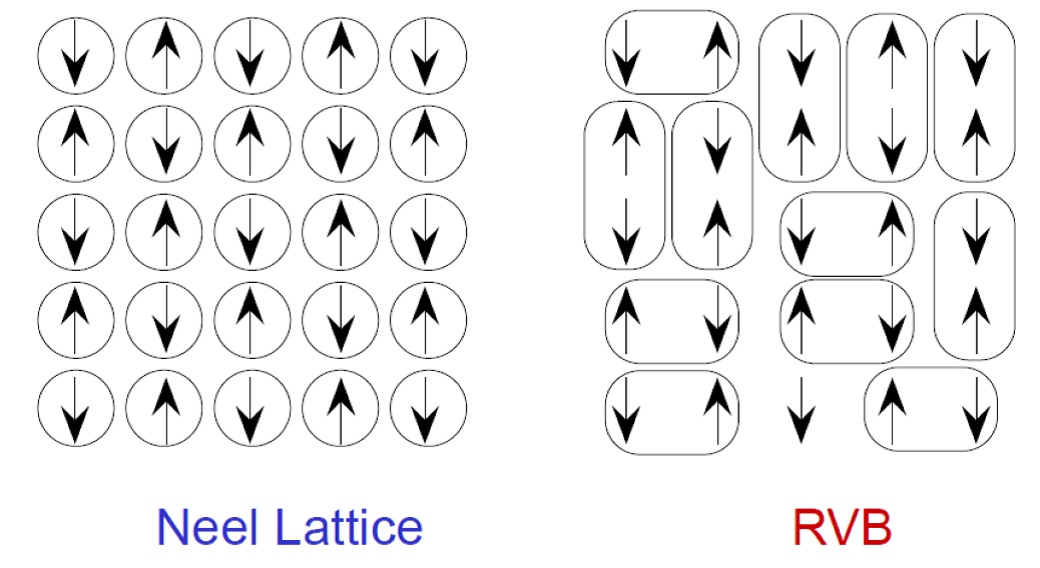}
\caption{In the figure is shown a "nematic" or random arrangement of electronic "directed" dimers in RVB at T*.
The Anderson resonating valence bond theory uses nearest neighbor two-qubit-singlet-entanglement Bell basis state $\Psi^{-}$ Hence, RVB theoretical framework can be considered as based on entanglement in the form of dimers.
[Reproduced from Ref. \cite{norman}]} \label{RVBdiag}
\end{figure}

\begin{figure}
[ht!]\centering\includegraphics[width=5.621300in]{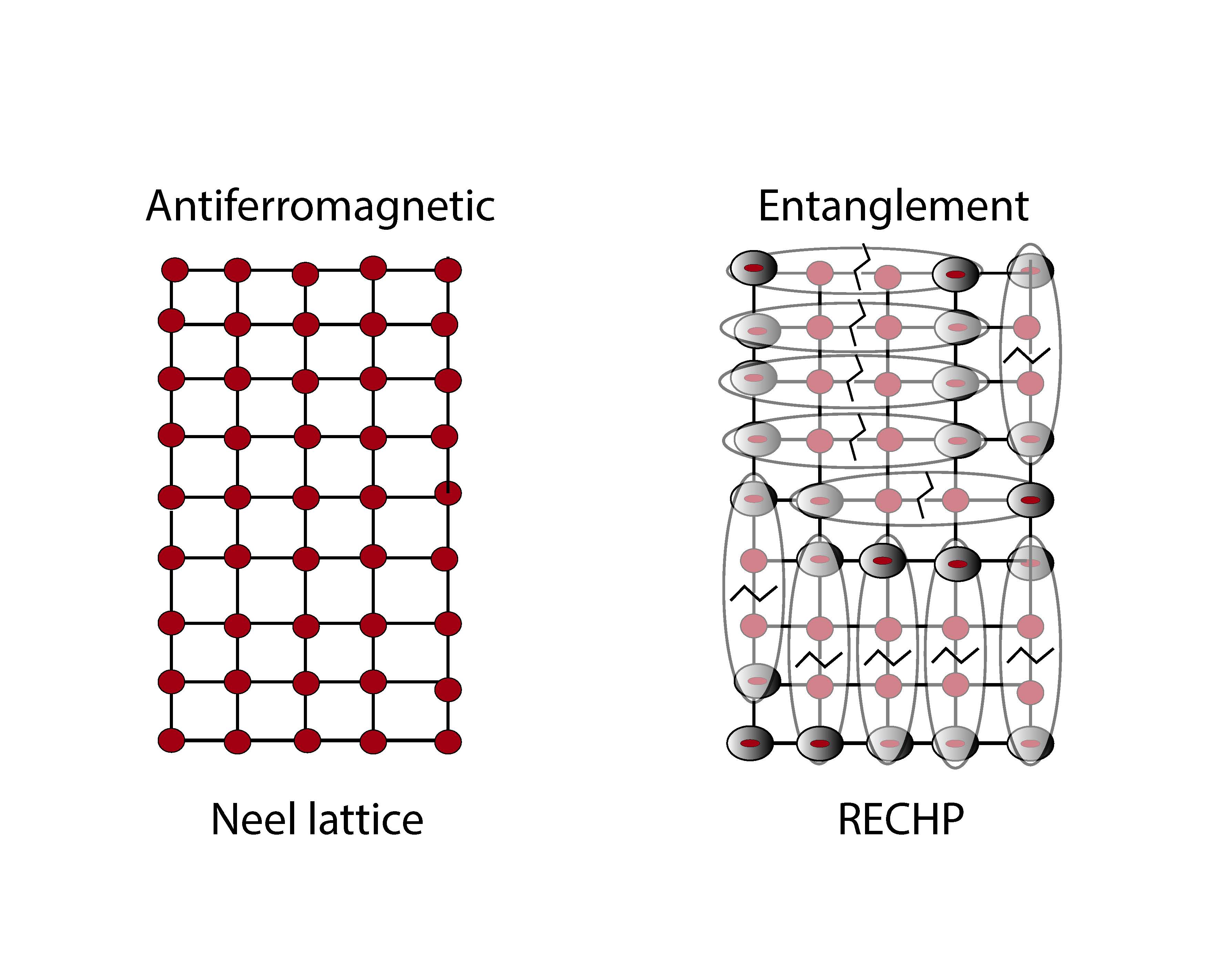}
\caption{The new RECHP theory of cuprates at the pseudogap phase, which resembles a "nematic
liquid" of extended entangled condensed pairs at T*. The spin arrows are suppressed in the figure. The short wiggly lines across antiferromagnetic links
indicate arbitrary operative length of the link dependent on the hole doping levels. We surmised that the
pseudogap phase with longer-chain link in the underdoped region is harder to transition to "smectic" order
than shorter chain link or the rate of transition is negligibly small for longer antiferromagnetic-chain link of
preformed pairs. This explains the gap between antiferromagnetic phase and the superconducting dome
of the phase diagram. We refer to this gap as the spin gap.} \label{RECPdiag}
\end{figure}

It is worth mentioning that aside from the coupling strength, there are
commonalities and as well as significant other differrences between the RVB
and RECHP. Both theories were based on entanglement pairing. Indeed, the
quantum state of RVB is assigned to one of the \textit{entanglement basis
states}, known as the \textit{Bell basis states} for bipartite entanglement
\cite{implement, buottele}, namely the singlet state $\Psi^{-}$%
\cite{anderson3}, as shown in Fig. \ref{RVBdiag} in the pseudogap disordered
state \cite{norman}. On the other hand, RECHP is based on the
\textit{doping-dependent} long-distance entanglement via the
antiferromagnetic-chain link, of \textit{doped holes }in Cu sites for
hole-doping (or \textit{generated} holes in oxygen sites for electron doping)
where the quantum states include both the singlet, $\Psi^{+}$, and triplet,
$\Phi^{+}$, Bell basis states, as illustrated in Fig. \ref{RECPdiag}. The use
of the term \textquotedblleft triplet\textquotedblright\ is actually a
misnomer here because the entangled system is not free to assume a singlet or
zero spin state. It has only two states of \textit{an emergent qubit}. Thus,
this term was used only as a label. Indeed, the transformation function
between triplet $\Phi^{\pm}$ and singlet $\Psi^{\pm}$ is the Pauli spin matrix
operator, $\sigma_{x}$. Of course, in Figs. \ref{RVBdiag} and \ref{RECPdiag},
spinons (\textit{unpaired electrons}) and holons (\textit{unpaired holes}) may
exist in very small numbers as imperfections and we ignore these "defects"
because these do not affect the essential physics that we focus on in this
paper. Here, we use freely the term "spinons" and "holons" not in the usual
sense of ground-state excitations of a many-body system.

Indeed, this new pairing mechanism in the \textit{configurational}
\textit{order }(CO)\textit{ zero-entropy phase} is characterized by parallel
superconducting stripes in the $CuO_{2}$ 2D plane, or rivers of charge
\cite{ref1, nanjing, ref2, ref3, ref4, ref5, ref6, ref7, ref8}. This occurs at
the SC dome of the phase diagram. These stripes also act as domain walls for
the antiferromagnetic order \cite{ref12, ref13, ref14}. The
\textit{resonating} concept of Anderson's RVB theory comes in the form of
equality relations between the EEF of \ different antiferromagnetic-chain link
configurations, for example, resonating between vertical and horizontal
"directed" extended-link configurations. In what follows, we appeal to
experiments and mean-field theory to place our ECHP mechanism on fundamental ground.

\subsection{Experimental validation of ECHP mechanism}

Here, we emphasize at the outset that realistic experimental works do exist
\cite{sahling} that validate the foundation of our proposed
entanglement-confinement pairing via an antiferromagnetic-chain link. The
experimental realization of antiferromagnetic-chain mediated entanglement
between distant spins in antiferromagnetic quantum spin chains was reported by
Shaling, et. al., using magnetic susceptibility and specific heat measurements
\cite{sahling}. In its most elementary viewpoint, one can visualize that in
any segment of an antiferromagnetic chain, the spin qubit at each end of the
chain is entangled, since any change of spin state at one end almost
immediately determines the spin state on the other end.

The view that quantum entanglements do not need interaction is a hype. This
popular belief is not a universally-accepted truth in the scientific
community. Indeed, in quantum gravity theory, the Einstein-Rosen (ER) bridge
or wormhole connects two entangled black holes. Generally, the ER is a
theoretical wormhole that connect two separate points in spacetime allowing
for instantaneous travel. \ Thus, the well-known EPR paradox still hold great
significance in theoretical physics. Indeed, some well-known theorists,
notably J. Maldacena of IAS (Princeton), and L. Susskind of Stanford, propose
the symbolic equality, namely, ER=EPR, which basically resolves the presence
of interaction between entangled systems. Others proposed that "EPR paradox"
in lower dimensions will find "ER" or interaction when viewed in higher
dimensions. This view has several classical analogies, such as a bound
vortex-antivortex pair on 2-D surface of swimming pool which has interaction
as a series \ of votices or "wormhole" hidden underneath the surface when
viewed in 3-D.

The low-temperature magnetization and specific heat studies by Bayat
\cite{bayat}, Sahling \cite{sahling}, and Sivkov \cite{sivkov} on the
antiferromagnetic-entanglement link between distant spins serve as direct
experimental evidence of our proposed entanglement pairing mechanism, thus
lending strong experimental support \cite{mitra} on the foundation of our new
pairing mechanism. Moreover, besides this experimental support, we now appeal
to mean field theory to further place our ECHP mechanism on firm fundamental
ground. In the following section, we show that the ECHP mechanism yields the
well-known dispersion relation characteristic of the gap-spectrum in strong
pairing theories.

\section{\label{theory copy(1)}Mean-field formulation of gap spectrum of ECHP}

We present the mean-field theory of hole-doped high-$T_{C}$ superconductivity
in cuprates. The characteristic superconductivity behavior of cuprates is
mainly due to the holes at the node. The most complete and general
nonequilibrium quantum transport theory of the superconductivity of charge
carriers in metals has been given by one of the authors (FAB) in a series of
publications \cite{buotrivista,buotbook,buotbook2}.

We considered a two-band model of high-$T_{C}$ cuprates. Luo and G. H Miley
\cite{ARPES} obtained the Fermi surface and band structure of generic cuprate
superconductors from ARPES. The electron pocket resides at the antinode and
the hole pocket resides at the node of the $d_{x^{2}-y^{2}}$ symmetric state.
We focus on hole pocket (at the node) to formulate the Hamiltonian.

In what follows, we will treat the main contributors to high-$T_{C}$
superconductivity in cuprates as holes \cite{egenerateh}. We write the
entangled states in terms of the hole-particle wavefunctions at positions $r$
and $r^{\prime}$ where $r\neq r^{\prime}$ for entangled distant pairs. We have
the \textit{complete Bell basis states} for the bipartite system, as follows
\cite{implement},%
\begin{equation}
\Phi^{+}=\frac{1}{\sqrt{2}}\left(  \varphi_{\downarrow}\left(  r\right)
\varphi_{\downarrow}\left(  r^{\prime}\right)  +\varphi_{\uparrow}\left(
r\right)  \varphi_{\uparrow}\left(  r^{\prime}\right)  \right)  ,\text{
\ }\Phi^{-}=\frac{1}{\sqrt{2}}\left(  \varphi_{\downarrow}\left(  r\right)
\varphi_{\downarrow}\left(  r^{\prime}\right)  -\varphi_{\uparrow}\left(
r\right)  \varphi_{\uparrow}\left(  r^{\prime}\right)  \right)  \label{bell1}%
\end{equation}%
\begin{equation}
\Psi^{+}=\frac{1}{\sqrt{2}}\left(  \varphi_{\downarrow}\left(  r\right)
\varphi_{\uparrow}\left(  r^{\prime}\right)  +\varphi_{\uparrow}\left(
r\right)  \varphi_{\downarrow}\left(  r^{\prime}\right)  \right)  \text{,
\ }\Psi^{-}=\frac{1}{\sqrt{2}}\left(  \varphi_{\downarrow}\left(  r\right)
\varphi_{\uparrow}\left(  r^{\prime}\right)  -\varphi_{\uparrow}\left(
r\right)  \varphi_{\downarrow}\left(  r^{\prime}\right)  \right)
\label{bell2}%
\end{equation}
where the $\varphi_{\sigma}$are the spin states often written in Dirac ket and
bra symbols, say $\left\vert \uparrow\right\rangle $, etc. These are related
to the spin field operators, $S_{x}$, $S_{y}$, and $S_{z}$:%
\begin{equation}
S_{x}=\hbar\Psi^{+}\text{, \ }S_{y}=i\hbar\Psi^{-}\text{ , \ }S_{z}=\hbar
\Phi^{-} \label{spin_op}%
\end{equation}
where the $\varphi_{\sigma}$ are now considered field operators in Eq.
\ref{spin_op}

For each hole pair, the corresponding entanglement states are generally placed
into the corresponding $\Phi^{\pm}$ and/or $\Psi^{\pm}$ states, where
$\Phi^{\pm}$ and $\Psi^{\pm}$ are the Bell basis states.The Hilbert-space
states are of the form,%
\begin{equation}
\Psi=%
{\displaystyle\sum\limits_{i\neq j}^{N}}
\left[
\begin{array}
[c]{c}%
C\left\{  \Phi_{ij}^{+}\Phi_{i^{\prime}j^{\prime}}^{+}...\Psi_{ij}^{+}%
\Psi_{i^{\prime}j^{\prime}}^{+}..\right\}
\end{array}
\right]  \label{hilbert}%
\end{equation}
In superconductivity pairing, we naturally consider only the $"+"$ Bell basis
states because $-J$ restricts the physics. In RVB theory, only the $\Psi^{-}%
$singlet Bell basis state was used \cite{anderson3}. It is difficult to
justify the negative term in $\Psi^{-}$ since $-J$ appears as a coefficient
for the antiferromagnetic domain. We expect this to be $\Psi^{+}$ Bell basis state.

For simplicity, we assume that for a given doping level, the effective
chain-length, $L_{eff}$, of all the entanglement species is equal. This
implies that the parameter $\Delta_{T^{\ast}}$ of the preformed pairing in the
underdoped region is a decreasing linear function of the doping levels, $d$,
because $L_{eff}$ decreases with increase of doping. Therefore, we have%
\begin{equation}
\Delta_{T^{\ast}}\equiv\Delta_{T^{\ast}}\left(  L_{eff}\left(  d\right)
\right)  \label{deltadopdep}%
\end{equation}
where $\Delta_{T^{\ast}}$increases with $L_{eff}\left(  d\right)  $ owing to
confinement. Moreover, $\Delta_{T^{\ast}}$ and $T^{\ast}$ are linear functions
of the doping levels, $d$. We emphasize that this is due to the concept of
hole confinement (or increasing $\Delta_{T^{\ast}}$ with length $L_{eff}$
which is conceptually nonexistent in the RVB nearest-neighbor local-bonding theory).

\subsection{Hamiltonian of node hole pocket}

For the holes at the node pocket of the Brillouin zone, we must consider the
kinetic energies of the holes at both ends of the antiferromagnetic-chain link
located at the unprime and prime coordinates. These holes act as 1D charge
carriers resembling the Fermi surface of metals in conventional BCS
superconductivity theory. The many-body spin field operator is defined,
\textit{\'{a}la} Schwinger \cite{thirdQ}, as%
\begin{equation}
\vec{S}\left(  \vec{r}\right)  =\frac{\hbar}{2}\psi_{\sigma}^{\dagger}\left(
\vec{r}\right)  \ \vec{\tau}_{\sigma\sigma^{\prime}}\ \psi_{\sigma^{\prime}%
}\left(  \vec{r}\right)  \label{schwinger}%
\end{equation}
In quantum-field representation,
\begin{align}
H_{node}  &  =\frac{1}{2}%
{\displaystyle\int}
dx\varphi^{\dagger}\left(  x\right)  \left(  -\frac{\hbar^{2}}{2m}\nabla
^{2}-\mu\right)  \varphi\left(  x\right)  +\frac{1}{2}%
{\displaystyle\int}
dy\varphi^{\dagger}\left(  y\right)  \left(  -\frac{\hbar^{2}}{2m}\nabla
^{2}-\mu\right)  \varphi\left(  y\right) \nonumber\\
&  -J\frac{1}{2}%
{\displaystyle\iint}
dxdy\varphi^{\dagger}\left(  x\right)  \varphi^{\dagger}\left(  y\right)
\vec{\tau}\left(  x\right)  \otimes_{ent}\vec{\tau}\left(  y\right)
\varphi\left(  y\right)  \varphi\left(  x\right) \label{ham1}\\
&  =\frac{1}{2}%
{\displaystyle\int}
dx\varphi^{\dagger}\left(  x\right)  \left(  E_{\nabla}^{\left(  h\right)
}-\mu\right)  \varphi\left(  x\right)  +\frac{1}{2}%
{\displaystyle\int}
dy\varphi^{\dagger}\left(  y\right)  \left(  E_{\nabla}^{\left(  h\right)
}-\mu\right)  \varphi\left(  y\right) \nonumber\\
&  -J\frac{1}{2}%
{\displaystyle\iint}
dxdy\varphi^{\dagger}\left(  x\right)  \varphi^{\dagger}\left(  y\right)
\vec{\tau}\left(  x\right)  \otimes_{ent}\vec{\tau}\left(  y\right)
\varphi\left(  y\right)  \varphi\left(  x\right)  \label{ham2}%
\end{align}
where $\otimes_{ent}$correspond to the l\textit{ong-range}
antiferromagnatic-mediated entanglement pairing of holes, $J$ is the
superexchange bonding, $\varphi$ is the hole quantum-field operator,
$E_{\nabla}^{\left(  h\right)  }$ is the differential kinetic operator for
holes corresponding to the parabolic band structure near the top of the hole
band, and $\vec{\tau}$ is the Pauli spin matrix operator. We can also write
the field-theoretical full Hamiltonian as,%
\begin{align}
H  &  =H_{1}+H_{2}+H_{ent}\nonumber\\
&  =\frac{1}{2}%
{\displaystyle\int}
dx\varphi^{\dagger}\left(  x\right)  \left(  -\frac{\hbar^{2}}{2m}\nabla
^{2}-\mu\right)  \varphi\left(  x\right)  +\frac{1}{2}%
{\displaystyle\iint}
dx\ dy\ \varphi^{\dagger}\left(  x\right)  \varphi^{\dagger}\left(  y\right)
V\left(  x,y\right)  \varphi\left(  y\right)  \varphi\left(  x\right)
\nonumber\\
&  -\frac{1}{2}\sum_{\sigma\sigma^{\prime}}%
{\displaystyle\iint}
dxdy\varphi_{\sigma}^{\dagger}\left(  x\right)  \varphi_{\sigma^{\prime}%
}^{\dagger}\left(  y\right)  \mathfrak{V}_{\left\vert x-y\right\vert }%
\varphi_{\sigma^{\prime}}\left(  y\right)  \varphi_{\sigma}\left(  x\right)
\label{totalH}%
\end{align}
where, $\mathfrak{V}_{\left\vert x-y\right\vert }=J\ \vec{\tau}\left(
x\right)  \otimes_{ent}\vec{\tau}\left(  y\right)  $ is determined by the hole
entanglement between sites $x$ and $y$ and $\left\vert x-y\right\vert
=L_{eff}>a$ is the lattice constant, in other words, $\otimes_{ent}$described
the effective antiferromagnetic-chain link $\Delta_{T^{\ast}}$ of Eq.
(\ref{deltadopdep}).

In the following, we neglect the Coulomb interactions in Eq. (\ref{totalH}),
which is already accounted for by the Mott insulator state of
antiferromagnetic chain, in a form of Coulomb blockade disallowing charge
transfer along the chain. However, the Coulomb scattering rates are later
included in the strange-metal-phase linear-$T$ resistivity. We obtain the
mean-field approximation of the last term as:%
\begin{align}
&  -\frac{1}{2}\sum_{\sigma\sigma^{\prime}}%
{\displaystyle\iint}
dxdy\varphi_{\sigma}^{\dagger}\left(  x\right)  \varphi_{\sigma^{\prime}%
}^{\dagger}\left(  y\right)  \mathfrak{V}_{\left\vert x-y\right\vert }%
\varphi_{\sigma^{\prime}}\left(  y\right)  \varphi_{\sigma}\left(  x\right)
\nonumber\\
&  \simeq-\frac{1}{2}\sum_{\sigma\sigma^{\prime}}%
{\displaystyle\iint}
dxdy\left\{  \left\langle \varphi_{\sigma}^{\dagger}\left(  x\right)
\varphi_{\sigma^{\prime}}^{\dagger}\left(  y\right)  \mathfrak{V}_{\left\vert
x-y\right\vert }\right\rangle \right\}  \varphi_{\sigma^{\prime}}\left(
y\right)  \varphi_{\sigma}\left(  x\right) \nonumber\\
&  -\frac{1}{2}\sum_{\sigma\sigma^{\prime}}%
{\displaystyle\iint}
dxdy\varphi_{\sigma}^{\dagger}\left(  x\right)  \varphi_{\sigma^{\prime}%
}^{\dagger}\left(  y\right)  \left\{  \left\langle \mathfrak{V}_{\left\vert
x-y\right\vert }\varphi_{\sigma^{\prime}}\left(  y\right)  \varphi_{\sigma
}\left(  x\right)  \right\rangle \right\}  \label{last term}%
\end{align}
and denote by
\begin{align}
\Delta^{\ast}\left(  L\right)   &  =\left\langle \varphi_{\sigma}^{\dagger
}\left(  x\right)  \varphi_{\sigma^{\prime}}^{\dagger}\left(  y\right)
\mathfrak{V}_{\left\vert x-y\right\vert }\right\rangle =J\times\ S\left(
L_{eff}\right)  _{\sigma\sigma^{\prime}}\label{pairpot1}\\
\Delta\left(  L\right)   &  =\left\langle \mathfrak{V}_{\left\vert
x-y\right\vert }\varphi_{\sigma^{\prime}}\left(  y\right)  \varphi_{\sigma
}\left(  x\right)  \right\rangle =J\times\ S\left(  L_{eff}\right)
_{\sigma^{\prime}\sigma} \label{pairpot2}%
\end{align}
where $S_{\alpha\beta}\left(  L_{eff}\right)  $ is the EEF of each pair at a
given doping level. We assumed that the effective entanglement
(antiferromagnetic)-chain length $L_{eff}$ is the same for a given hole-doping
level, that is, $L_{eff}$ is only a single-valued function of the doping
concentration. For a given antiferromagnetic chain length, either triplet or
singlet pairing has the same average value for EEF or $S_{\alpha\beta}\left(
L_{eff}\right)  $. In terms of bond terminology, on average we assume the same
number of bonds for triplet or singlet pairings. This is demonstrated in Fig.
\ref{figchain_reduce} for singlet pairing.

The second term of Eq. (\ref{last term}) can be written in the mean field
approximation as:%
\begin{align}
&  -J\frac{1}{2}%
{\displaystyle\iint}
dxdy\varphi^{\dagger}\left(  x\right)  \varphi^{\dagger}\left(  y\right)
\tau\left(  x\right)  \otimes_{ent}\tau\left(  y\right)  \varphi\left(
y\right)  \varphi\left(  x\right) \nonumber\\
&  =-J\frac{1}{2}%
{\displaystyle\iint}
dxdy\ \Delta_{h}^{\dagger}\left(  x,y\right)  \ \varphi\left(  y\right)
\varphi\left(  x\right)  -J\frac{1}{2}%
{\displaystyle\iint}
dxdy\ \Delta_{h}\left(  x,y\right)  \ \varphi^{\dagger}\left(  y\right)
\varphi^{\dagger}\left(  x\right)  \label{secondterm}%
\end{align}

We have for the pairing terms, accounting for spins,%
\begin{align}
-J\frac{1}{2}\varphi\left(  r\right)  _{\sigma}\varphi\left(  r^{\prime
}\right)  _{\sigma^{\prime}}  &  =-J\frac{1}{2}\left[  \varphi\left(
r\right)  _{\downarrow}\varphi\left(  r^{\prime}\right)  _{\downarrow}%
+\varphi\left(  r\right)  _{\downarrow}\varphi\left(  r^{\prime}\right)
_{\uparrow}+\varphi\left(  r\right)  _{\uparrow}\varphi\left(  r^{\prime
}\right)  _{\downarrow}+\varphi\left(  r\right)  _{\uparrow}\varphi\left(
r^{\prime}\right)  _{\uparrow}\right] \nonumber\\
&  \Longrightarrow-J\frac{1}{\sqrt{2}}\left[  \Phi^{+}\left(  r,r^{\prime
}\right)  _{\sigma\sigma^{\prime}}+\Psi^{+}\left(  r,r^{\prime}\right)
_{\sigma\sigma^{\prime}}\right]  \label{bellbasispair}%
\end{align}
where the second line explicitly shows the Bell basis states
\cite{perspective}. Similarly, we have,%
\begin{align}
-J\frac{1}{2}\varphi^{\dagger}\left(  r\right)  _{\sigma}\varphi^{\dagger
}\left(  r^{\prime}\right)  _{\sigma^{\prime}}  &  =-J\frac{1}{2}\left[
\varphi^{\dagger}\left(  r\right)  _{\downarrow}\varphi^{\dagger}\left(
r^{\prime}\right)  _{\downarrow}+\varphi^{\dagger}\left(  r\right)
_{\downarrow}\varphi^{\dagger}\left(  r^{\prime}\right)  _{\uparrow}%
+\varphi^{\dagger}\left(  r\right)  _{\uparrow}\varphi^{\dagger}\left(
r^{\prime}\right)  _{\downarrow}+\varphi^{\dagger}\left(  r\right)
_{\uparrow}\varphi^{\dagger}\left(  r^{\prime}\right)  _{\uparrow}\right]
\nonumber\\
&  \Longrightarrow-J\frac{1}{\sqrt{2}}\left[  \Phi^{+}\left(  r,r^{\prime
}\right)  _{\sigma\sigma^{\prime}}^{\ast}+\Psi^{+}\left(  r,r^{\prime}\right)
_{\sigma\sigma^{\prime}}^{\ast}\right]  \label{entTermH}%
\end{align}
clearly showing the mathematical expression for the sum of triplet and singlet
hole pairings.

\subsection{Equation of motion for entangled holes for chain length $L_{eff}$}

We explicitly exhibit the spin index, as well as the coordinates of the holes
on both ends of an entangled pair, in the form of a column vector $\left(
note\ r\neq r^{\prime}\text{, }\left\vert r-r^{\prime}\right\vert =L>a\right)
$, where the equation of motion is given by,%
\begin{equation}
i\hbar\frac{\partial}{\partial t}\Phi=\left[  \Phi,H\right]  =M\ \Phi
\label{eqmotion}%
\end{equation}
Writing explicitly, this is,%

\begin{equation}
i\hbar\frac{\partial}{\partial t}\left(
\begin{array}
[c]{c}%
\varphi\left(  r\right)  _{\downarrow}\\
\varphi\left(  r\right)  _{\uparrow}\\
\varphi\left(  r^{\prime}\right)  _{\downarrow}\\
\varphi\left(  r^{\prime}\right)  _{\uparrow}\\
\varphi^{\dagger}\left(  r\right)  _{\downarrow}\\
\varphi^{\dagger}\left(  r\right)  _{\uparrow}\\
\varphi^{\dagger}\left(  r^{\prime}\right)  _{\downarrow}\\
\varphi^{\dagger}\left(  r^{\prime}\right)  _{\uparrow}%
\end{array}
\right)  =M\times\left(
\begin{array}
[c]{c}%
\varphi\left(  r\right)  _{\downarrow}\\
\varphi\left(  r\right)  _{\uparrow}\\
\varphi\left(  r^{\prime}\right)  _{\downarrow}\\
\varphi\left(  r^{\prime}\right)  _{\uparrow}\\
\varphi^{\dagger}\left(  r\right)  _{\downarrow}\\
\varphi^{\dagger}\left(  r\right)  _{\uparrow}\\
\varphi^{\dagger}\left(  r^{\prime}\right)  _{\downarrow}\\
\varphi^{\dagger}\left(  r^{\prime}\right)  _{\uparrow}%
\end{array}
\right)  \label{matrixevolve}%
\end{equation}
where $M$ denotess an $8\times8$ matrix. We expect that the equation for
$\varphi\left(  r\right)  _{\sigma}$ will be coupled to $\varphi^{\dagger
}\left(  r^{\prime}\right)  _{\sigma^{\prime}}$ as expected for entangled pairs.

To evaluate the equation of motion, we evaluated with respect to the
entanglement term of the Hamiltonian given by Eq. (\ref{entTermH}). Let us
consider a unit hole pair of length $L_{eff}$ shown in Fig.
\ref{figchain_reduce} for a given doping concentration. Then we put
\begin{equation}
r^{\prime}=r+L_{eff}=z \label{coord}%
\end{equation}
Pursuing Eq. (\ref{matrixevolve}), we obtain

\bigskip%
\begin{align}
&  i\hbar\frac{\partial}{\partial t}\left(
\begin{array}
[c]{c}%
\varphi\left(  r\right)  _{\downarrow}\\
\varphi\left(  r\right)  _{\uparrow}\\
\varphi\left(  z\right)  _{\downarrow}\\
\varphi\left(  z\right)  _{\uparrow}\\
\varphi^{\dagger}\left(  r\right)  _{\downarrow}\\
\varphi^{\dagger}\left(  r\right)  _{\uparrow}\\
\varphi^{\dagger}\left(  z\right)  _{\downarrow}\\
\varphi^{\dagger}\left(  z\right)  _{\uparrow}%
\end{array}
\right) \nonumber\\
&  =\left(
\begin{array}
[c]{cccccccc}%
\left(  \varepsilon_{\varphi}\right)  & 0 & 0 & 0 & 0 & 0 & \Delta & \Delta\\
0 & \left(  \varepsilon_{\varphi}\right)  & 0 & 0 & 0 & 0 & \Delta & \Delta\\
0 & 0 & \left(  \varepsilon_{\varphi}\right)  & 0 & -\Delta & -\Delta & 0 &
9\\
0 & 0 & 0 & \left(  \varepsilon_{\varphi}\right)  & -\Delta & -\Delta & 0 &
0\\
0 & 0 & \Delta & \Delta & \left(  \varepsilon_{\varphi}\right)  & 0 & 0 & 0\\
0 & 0 & \Delta & \Delta & 0 & \left(  \varepsilon_{\varphi}\right)  & 0 & 0\\
-\Delta & -\Delta & 0 & 0 & 0 & 0 & \left(  \varepsilon_{\varphi}\right)  &
0\\
-\Delta & -\Delta & 0 & 0 & 0 & 0 & 0 & \left(  \varepsilon_{\varphi}\right)
\end{array}
\right) \nonumber\\
&  \times\left(
\begin{array}
[c]{c}%
\varphi\left(  r\right)  _{\downarrow}\\
\varphi\left(  r\right)  _{\uparrow}\\
\varphi\left(  z\right)  _{\downarrow}\\
\varphi\left(  z\right)  _{\uparrow}\\
\varphi^{\dagger}\left(  r\right)  _{\downarrow}\\
\varphi^{\dagger}\left(  r\right)  _{\uparrow}\\
\varphi^{\dagger}\left(  z\right)  _{\downarrow}\\
\varphi^{\dagger}\left(  z\right)  _{\uparrow}%
\end{array}
\right)  \label{matrixeq}%
\end{align}
where $\varepsilon_{\varphi}$ is the kinetic energy of the holes determined by
the band structure near the top of the valence band. The $\varepsilon
_{\varphi}$ value measured from the chemical potential, corresponds to the
holes residing in the stripy channel.

\subsubsection{Eigenvalue equation for holes}

The effective Schr\"{o}dinger equation in many-body quantum field theory
follows from the equation for the expectation or average of the quantum field
operator. Some instances of these are typefied by the Gross-Pitaevskii
equation for bosons and the Kohn--Sham equations or density functional
theories for fermions. Thus, we identify the equations for the wavefunctions
from the obtained quantum-field operator equations, Eq. (\ref{matrixeq}). The
resulting eigenvalue equation follows,%
\begin{equation}
\left\vert \left(
\begin{array}
[c]{cccccccc}%
\left(  \left[  \varepsilon_{\varphi}-\mu\right]  -E\right)  & 0 & 0 & 0 & 0 &
0 & \Delta & \Delta\\
0 & \left(  \varepsilon_{\varphi}-E\right)  & 0 & 0 & 0 & 0 & \Delta &
\Delta\\
0 & 0 & \left(  \varepsilon_{\varphi}-E\right)  & 0 & -\Delta & -\Delta & 0 &
0\\
0 & 0 & 0 & \left(  \varepsilon_{\varphi}-E\right)  & -\Delta & -\Delta & 0 &
0\\
0 & 0 & -\Delta & -\Delta & -\left(  \varepsilon_{\varphi}+E\right)  & 0 & 0 &
0\\
0 & 0 & -\Delta & -\Delta & 0 & -\left(  \varepsilon_{\varphi}+E\right)  & 0 &
0\\
\Delta & \Delta & 0 & 0 & 0 & 0 & -\left(  \varepsilon_{\varphi}+E\right)  &
0\\
\Delta & \Delta & 0 & 0 & 0 & 0 & 0 & -\left(  \varepsilon_{\varphi}+E\right)
\end{array}
\right)  \right\vert {\small =0} \label{eig}%
\end{equation}
The result is%
\begin{equation}
E=\pm\sqrt{\varepsilon_{\varphi}^{2}+\Delta^{2}} \label{gapeq}%
\end{equation}
which is a well-known dispersion relation for the gap-spectrum in strong
pairing theories.

\section{Entanglement and confinement in RECHP theory}

We proposed a drastic conceptual and physical reformulation of the RVB theory,
in which the essential ingredients are the \textit{long-range} entanglement
and confinement mechanisms (entanglement measure) between \textit{doped holes}
rather than the local nearest-neighbor entanglement of unpaired
\textit{electrons} in Cu sites, the so-called "directed dimer" states forming
the RVB theory. One might consider our theory to be a resonating
entanglement-confinement hole pair (RECHP) theory where confinement is an
additional crucial physical concept missing in local dimer or two-nearest
qubit entanglement of RVB theory.

\subsection{The pairing order and symmetry-breaking order}

Two different concepts of order must be recognized in high-$T_{C}$ cuprates.
The first is the more familiar form of pairing order (PO), that \ is, the
condensation of the pair to the lowest energy of its gap-spectrum, which
occurs at a temperature $T=T^{\ast}$. In contrast, CO \ is a type of SB phase
ordering, from "nematic" to "smectic," of the "directed"
antiferromagnetic-chain links. This produces a current-carrying CO phase, as
shown in Fig. \ref{fig3} where the periodic arrangement of alternate $\Phi$
and $\Psi$ entangled hole-pair segments is shown. CO may or may not occur in
the underdoped region after $T^{\ast}$ upon further cooling. If CO fails to
materialize upon cooling, the result is the so-called "spin gap" which occurrs
between the end of AF order and the rise of SC dome. We note that PO and CO
occur jointly at the peak of the SC dome as well as in the overdoped regions
of the SC dome. In the CO ordered phase, all these entangled pairs were
degenerate. The "bare" $\Phi$ and $\Psi$ series arrangement of Bell basis
states are thought to be the dominant contribution in the underdoped region of
cuprates, resulting in unpolarized current stripes in Fig. \ref{fig3}. In Fig.
\ref{fig4}, is shown where triplet entanglements \textit{acting as emergent
qubits} \cite{perspective,buotbook2} are entangled as singlet and vice versa
in a synthesis manner. This seems to be dominant in the overdoped regions in
special cases, that is, under local symmetry-breaking crystal distortions.
These views are in agreement with the results of SR-ARPES experiments. All of
these CO orders generate 'stripy' superconductivity patterns, as illustrated
in Figs. \ref{fig3} and \ref{fig4}.

\subsection{The path from PO to CO: the evolution in configurational entropy
(CE) space}

Since the spatial arrangement of preformed pairs are distributed randomly in
the vertical and horizontal directions, we may characterized this situation
with one variable, namely, the entropy. One can use Shannon entropy formula
for nonuniform random distribution or Boltzmann entropy for uniform
distribution of distinct random arrangements. For example, at a certain doping
level we may assume that the total distinct random arrangements in $x$- and
$y$- directions may be some large number $\Omega$. The configurational entropy
(CE) is then proportional to $Log\ \Omega$, this seems a convenient single
quantity to characterize the evolution of disordered spatial configurations of
preformed pairs as the temperature is further lowered from PO at $T^{\ast}$.
Clearly, zero $CE$ signifies that the spatial configurations of preformed
pairs undergo symmetry-breaking \textquotedblleft smectic\textquotedblright%
\ configurational transition to a definite unique rearrangement, and at CO we
have,
\begin{equation}
CE=Log\ \left(  1\right)  =0\label{logzero}%
\end{equation}
at the SC dome.

\subsection{The drive to CO: parallel conducting stripes}

\begin{figure}
[ht!]\centering\includegraphics[width=3.6927in]{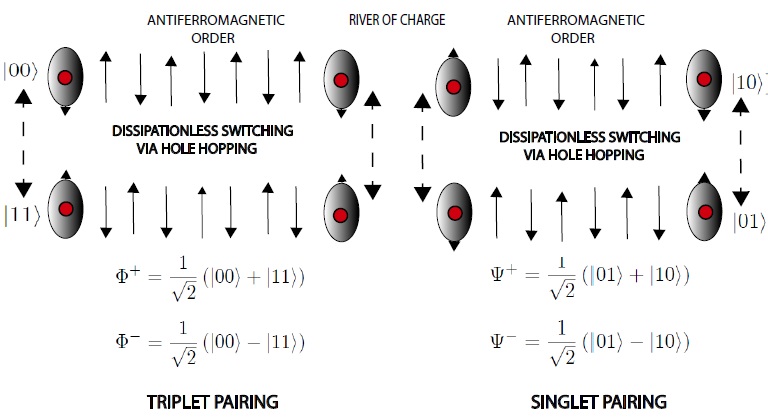}
\caption{In the CO "smectic" order (i.e., CE=0) phase of some materials, the corresponding Bell basis states are arranged as shown. These two sections of entangled hole pairs, namely, the $\Phi$ section and the $\Psi$ section is arranged into alternate sections, periodic "lattice" along horizontal $x$- and layered along $y$- directions to form current-carrying condensed pattern of entangled pairs and rivers of charge at their ends with zero net spins (unpolarized) as shown. The hole "blob' accounts for the complex dressing of the holes in response to the coupling with the background. \cite{weng,weng2,weng3, danilov, mottness}.}\label{fig3}%

\end{figure}

\begin{figure}
[ht!]\centering\includegraphics[width=4.2099in]{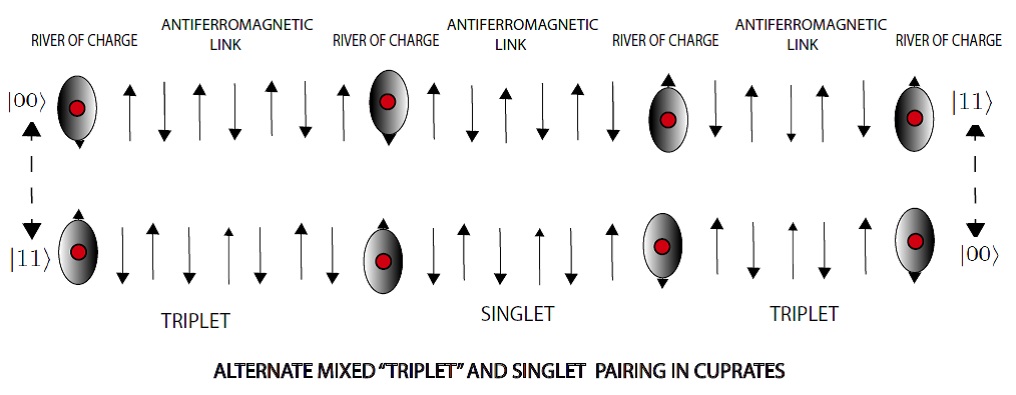}
\caption{The CO of antiferromagnetic links showing a synthesis or compounding of interacting triplet
and singlet entanglements. This can happen when some symmetry-breaking crystal distortion occurs especially
at overdoping. This gives spin-polarized SR-ARPES results and is deemed to be due to surface distortion in
overdoped regions. This is a higher-order entanglement process where triplet entanglements, by acting as
emergent qubits, are being entangled as singlet and vice versa. This is presumably dominant in the overdoped
region of the phase diagram of some materials.} \label{fig4}
\end{figure}

Disordered preformed RECHP in vertical and horizontal "directed" pairings,
without current flow, possess high "nematic" symmetry with resonating
horizontal and vertical antiferromagnetic links. This may be considered as
nurturing the capacitive energies in the system owing to the electrostatic
interactions between the directed pairings and background charges. At low
temperatures, SB transition from the "nematic" to the "smectic" order is a
zero CE-state configurational phase and therefore gives the minimum in the
energy landscape. The transition from "nematic" to "smectic" configuration is
a symmetry-breaking process. Thus, thermodynamic energy is released upon
cooling towards CO allowing the current to flow through the parallel 1D
conducting stripes, as illustrated in Figs. \ref{fig3} and \ref{fig4}.

The tendency to attain "smectic" CO depends on the
antiferromagnetic-chain-link length. It is infinitely slower (low resonance
frequency) to achieve the SB order for longer and more stable
antiferromagnetic-chain links than for shorter ones. This is described by the
rate of change of CE with respect to temperature, see Sec.\ref{rate} below.
This very slow rate leads to a spin gap in the underdoped regions of the
cuprates phase diagram at low doping levels, as shown in Fig. \ref{phaseD}.
The "smectic" CO once attained persist in the overdoped regions of the SC dome
and even for $T>T_{C}$, which turns out to be compatible with the linear-T
resistivity of the strange metal phase, discussed in Sec.\ref{rate}.

\subsubsection{Mode of superconduction in "condensed" PO and CO phase}

In the SC dome, the holes move at both ends of the
entanglement-antiferromagnetic-link in unison, effectively performing
dissipationless switching between the two layers, as illustrated in Fig.
\ref{fig3}. This serves as a virtual switching between the two qubit states of
the entangled holes. This also effectively result in a \textit{virtual}
superconduction of "extended boson" without dissipation, oscillating between
its two degenerate qubit eigenstates while being transported with charge
$2\left\vert e\right\vert $, where $e$ is the electron charge. This is the
nature of the current flow in our new long-range hole-pairing mechanism. We
are now ready to present the analysis of the entire phase diagram of cuprates.

\section{\label{rate}RECHP theory of the high-$T_{C}$ cuprates: analytical and
graphical analysis}

\begin{figure}
[ht!]%
\centering\includegraphics[width=5.849600in]{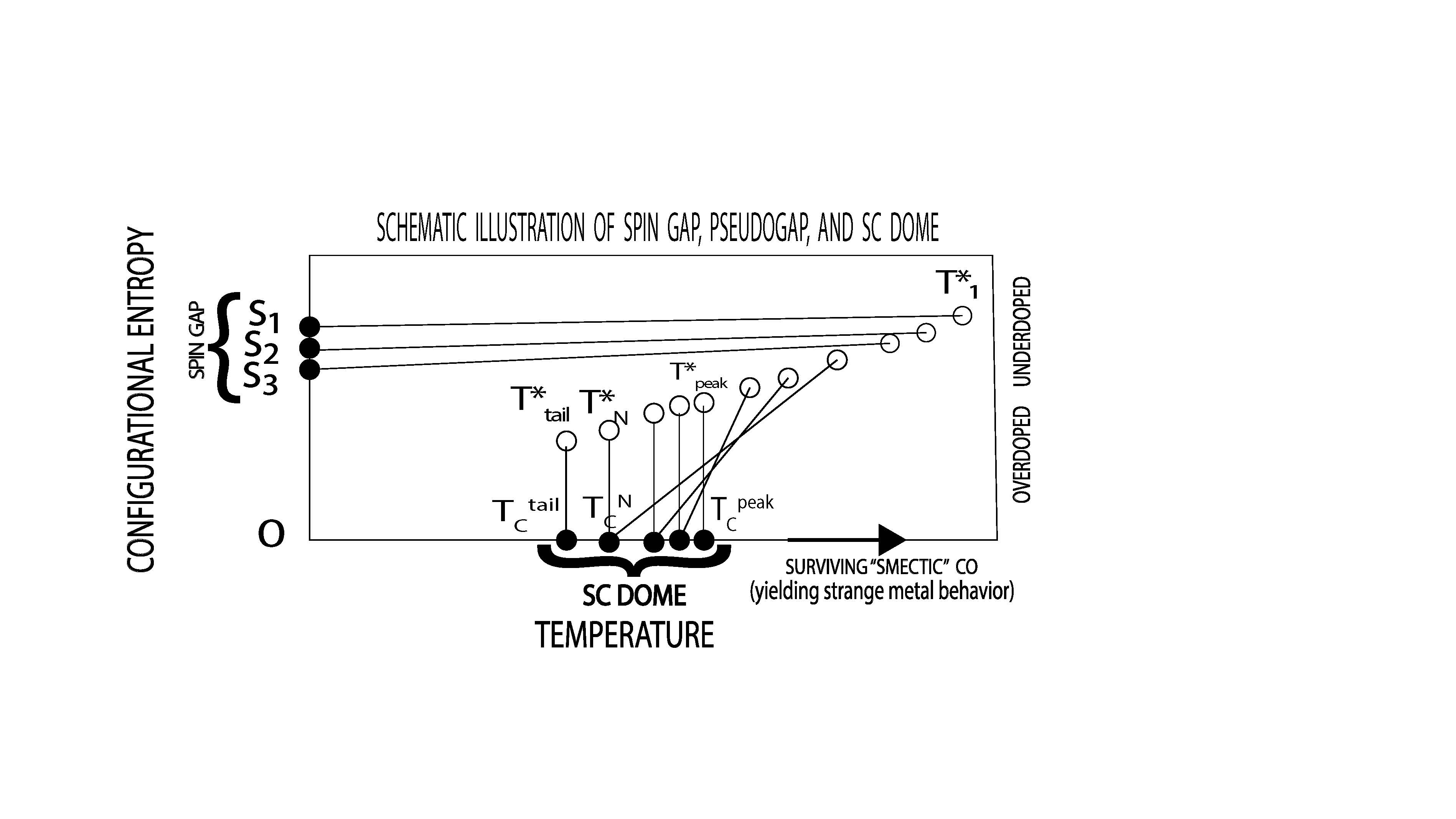}\caption{Graphical analysis of the process leading to spin gap, SC dome, and strange metal phase. The placements of $T^{\ast }$ in
the plot is compatible with decrease in $T^{\ast }$ with increase in doping and decrease in temperature. The CE vs. T relation is assume to obey linear transition from T* with the rate R increasing with increase of doping levels. The increase in R is consistent with the increase in the resonating frequency with decrease in the effective antiferromagnetic-chain link. The figure shows a singularity, where the rate R goes to infinity, manisfested as  a "kink" at the peak of the SC dome and down through the entire overdoped region of the SC dome. It retraces the locus (black dots) of the CO produced by T* in the underdoped region thus forming the SC dome. This singularity must be interpreted as the joint occurrence of $T^{\ast }$ and $T_{C}$, either as a coincident or tangent. Our graphical analysis clearly shows this as coincident $T^{\ast }$ and $T_{C}$  plots. The $T^{\ast }$ curve coincide with the $T_{C}$  curve, since no temperature change is needed to evolve from $T^{\ast }$ to $T_{C}$. The arrow denotes the surviving CO at temperature above SC dome, maintaining the one-dimensional stripes obeying the laws of Planckian mesoscopic physics. This yields a linear T-dependence of resistivity of the strange metal, discussed in the text.}\label{schematic}%

\end{figure}

Here, we give the RECHP theory of the entire phase diagram of high-$T_{C}$
cuprates.  The pseudogap phase is thus defined in terms of the condensation of
randomly "directed" pairings to their lower $E=-\sqrt{\varepsilon_{\varphi
}^{2}+\Delta^{2}}$ at temperature $T^{\ast}$, that is, to a low-energy phase
transition causing a loss in the density of states. In the pseudogap phase,
$\Delta=\Delta_{T^{\ast}}$ for disordered preformed pairs. \ In Fig.
\ref{schematic}, we graphically analyze the spin gap, pseudogap, and SC dome,
as well as the mechanism leading to strange metal behavior. It is worth
emphasizing that our new hole pairing mechanism predicts the spin gap, the
singularity of $T^{\ast}$at the SC dome peak, and the coincidence of $T^{\ast
}=T_{C}$ in the overdoped regions of the SC dome, graphically shown in Fig.
\ref{schematic} consistent with experiment in Fig. \ref{phaseD}. Other
theories predict the entire analyticity of $T^{\ast}$ being tangent to $T_{C}$
in the overdoped region. The entire analyticity of $T^{\ast}$ of other
theories does not agree with the experimental phase diagram.

The temperature dependence of configurational entropy $(CE)$ is simply
determined from the understanding that the onset of superconductivity or the
symmetry-breaking \textquotedblleft smectic\textquotedblright\ transition
occurs when the temperature is being lowered. Here we simply assume a linear
function of the temperature-dependence of $CE$ shown in Fig.\ref{schematic}.
The slope, $R$, of this linear function, i.e., lines ($CE\left(  T\right)
^{\prime}s$) emanating from each $T^{\ast}$-point, indicate the \textit{rate
of evolution} towards the superconductivity phase ($CE=0$) as the temperature
is lowered. This is given by%
\begin{equation}
R=\frac{\partial CE\left(  T\right)  }{\partial T}\label{rigid}%
\end{equation}
The rate $R$ is thus a measure of the \textquotedblleft
flexibility\textquotedblright\ of the random arrangements of preforemed pairss
to evolve towards symmetry-breaking \textquotedblleft
smectic\textquotedblright\ CO phase transition where,%
\begin{equation}
CE=\log\left(  1\right)  =0\label{zeroCE}%
\end{equation}
at the SC dome. This, $R\ll\eta,$ is very small in the spin-gap region of the
doping levels in the phase diagram. Moreover, we emphasize that this slope,
$R$, is proportional to the doping level since from the experimental phase
diagram, the spin gap serve as evidence that entanglements with longer
antiferromagnetic-chain link, which happens at smaller doping levels, gives
much smaller $R$, i.e., it is harder for the $CE\left(  T\right)  $ to change
with corresponding change of temperature. We therefore take $R$ to be
proportional to the doping levels, $d$,%
\begin{equation}
R=\omega\ d.\label{propw}%
\end{equation}
where $\omega$ is the proportionality constant in the underdoped region but
blows up at the peak of the SC dome . In other words, entanglements with
shorter antiferromagnetic-chain link oscillate at higher frequency than for
entanglements with longer antiferromagnetic-chain link. In Fig.
\ref{schematic}, is shown the singularity in $R$ at the peak of the SC dome
and all through the overdoped region of the dome. This signifies that the
$T^{\ast}$ curve coincide with the $T_{C}$ curve in the overdoped region,
since no change in temperature occurs in going from $T^{\ast}$ to $T_{C}$ as
graphically depicted in Fig. \ref{schematic}

From the viewpoint of our new hole-pairing perspective, we analyze in detail
the typical phase diagram (Fig. \ref{phaseD}) of high-T$_{C}$ cuprates as follows.

\subsection{Pseudogap in underdoped region}

The pseudogap region probably initially involves the motion of a single hole
upon dilute doping in the antiferromagnetic domain \cite{sachdev, ref10,
henning}. Entanglement pairing immediately starts as soon as sufficient holes
are introduced concomitant with the destruction of the antiferromagnetic
long-range order; however, at much smaller doping (with doped holes occupying
$d_{x^{2}-y^{2}}$ at the copper sites), the antiferromagnetic order may still
be supported but at a rapidly decreasing Neel temperature, $T_{N}$. The
pseudogap phase in the underdoped cuprates involves the PO condensation of
randomly distributed "directed" preformed pairs along $x$- and $y$-directions
of the $CuO_{2}$ plane to their lower energies of their gap-spectrum upon
cooling to $T^{\ast}$. This occurs whenever electrons \cite{e as h,egenerateh}
or holes are introduced into a magnetically insulating antiferromagnetic
domain forming a "nematic"-like disordered "directed" pairings
\cite{nematic,levin}. All pairings contribute to lower energies upon cooling
to $T=T^{\ast}$as suggested by Eq. (\ref{gapeq}).

\subsubsection{Confinement effects}

In the pseudogap of the underdoped regions, the confinement effects predict
decreasing $T^{\ast}$ with increasing hole-doping concentration levels, $d$,
with $\Delta_{T^{\ast}}\Longrightarrow\Delta_{C}$ at the optimum doping or
peak of the SC dome, as shown in Fig. \ref{schematic}. We have%
\begin{equation}
\Delta^{\ast}\left(  S_{ent}\left(  L_{eff}\right)  \right)  =J\times
S_{ent}\left(  L_{eff}\right)  =\beta L_{eff} \label{beta}%
\end{equation}
where $S_{ent}\left(  L_{eff}\right)  $ is the EEF as a function of the
chain-entanglement-link-length, $L_{eff}$. Here $\beta=\frac{J}{L_{B}}$, where
$L_{B}$ is the bond length (nearest neighbor Cu site-Cu site distance).

We believe that some experiments on the fluctuating magnetic order carry
different length scales, the so-called correlation length, compared to the
entangled pair antiferromagnetic-link length, where signals are propagated
from one end to the other entangled end of the chain. As experimentally probed
by Shaling et al \cite{sahling} on antiferromagnetic chain entanglement, they
used magnetic susceptibility and heat capacity measurements which generally
prove long-distance entanglement and therefore expected to have different
length scales than those obtained from the fluctuating magnetic order obtained
from neutron scattering experiments \cite{neutron}.

Using the effective length as a decreasing function of doping, we have,
\begin{equation}
\frac{\partial L_{eff}}{\partial d}=-\gamma\label{proport1}%
\end{equation}%
\begin{equation}
L_{eff}=L_{o}-\gamma d \label{proport2}%
\end{equation}
where $\gamma$ is the absolute value of the slope of the linear dependence of
$T^{\ast}$ at the pseudogap. Therefore,%
\begin{align}
\Delta^{\ast}\left(  S_{ent}\left(  L_{eff}\right)  \right)   &  =\beta\left(
L_{o}-\gamma d\right) \label{delta1}\\
T^{\ast}  &  =\kappa\left(  L_{o}-\gamma d\right)  \label{Taster}%
\end{align}
The "spin gap" between the complete disappearance of antiferromagnetism and
the rise of SC dome in Fig. \ref{phaseD} is a mark of disordered "directed"
very long pairings that does not CO into a "smectic" configuration with a
continued decrease in temperature, so as to release capacitive energies by
allowing current flow through a pattern of stripes, (refer to Fig.
\ref{schematic}). The pseudogap region consists of phase-disordered entangled
pairs with increasing $R$ trends with hole doping, that is with a decrease in
the \textit{effective or average} length of the
antiferromagnetic-chain-entanglement link as the doping level increases. This
results in a linear $\Delta_{T^{\ast}}$ with doping, as expressed in Eq.
(\ref{delta1}). \ We emphasize that because of the singularity of $R,$this
manifest in the non-analyticity of $T^{\ast}$, Eq. (\ref{Taster}), turning
eventually into a "kink" at the SC dome peak. This is a crucial point in
agreement with experimental phase diagram of Fig. \ref{phaseD}.

\subsection{Graphical analysis of the SC dome and the $T^{\ast}$ singularity}

As mentioned before, confinement effects predict decreasing $T^{\ast}$ with
increasing hole doping levels, $d$, with important results that $\Delta
_{T^{\ast}}\Longrightarrow\Delta_{C}$ at the optimum doping or peak of the SC
dome. This is portrayed in Fig. \ref{schematic} as an $R$ \textit{singularity}
of the $T^{\ast}$-linear curve producing a "kink" at the SC dome peak. The
\textit{competition} between decrease in antiferromagnetic chain-link
entanglement versus increase in $R$ (or \textit{resonant frequency increase
with decrease in }$\Delta_{T^{\ast}}$) culminates in \textit{singularity} of
the slope of $T^{\ast}$-linear curve at the peak of the SC dome where
$\Delta_{T^{\ast}}\Longrightarrow\Delta_{C}$, continuing into the overdoped
regions of the SC dome. This means that the superconducting gap, $\Delta_{C}$,
and pseudogap, $\Delta_{T^{\ast}}$, are coincident (as opposed to tangent in
some literature) in the overdoped region of the SC dome, as indicated in Fig.
\ref{schematic}. In addition, the temperature $T^{\ast}=T_{C}$ in the
overdoped region of the dome. The entanglement and confinement of holes
becomes phase coherent, in the sense of symmetry-breaking "smectic" CO which
is carrying current with respect to the in plane electric fields forming
rivers of charge in a parallel stripy pattern in the SC dome region. At
optimal doping, the main contribution comes from the ordered pattern of
entangled holes as depicted in Fig. \ref{fig3}, which form parallel rivers of
charge. In the overdoped region, higher-order contributions with
\textit{spin-polarized stripes,} illustrated in Fig. \ref{fig4}, may
contribute significantly in special cases, as indicated by experiments
\cite{lou, iwasawa,gotlieb}.

\subsection{Strange metal\ phase: 1D stripes and linear-T resistivity}

Above $T_{C}$, the entangled pairs are no longer degenerate, but the CO still
survives above $T_{C}$ \cite{huang}; therefore, preserving the parallel
conducting stripes, each stripe can be approximated as one-dimensional. This
is illustrated by an arrow in Fig. \ref{schematic}. We show that the
resistivity is linear compared with that of conventional metals \cite{ref15}.

Indeed, the universal conductivity in the strange metal region of the phase
diagram indicates one-dimensional quantum transport, leading to the Planckian
conductivity \cite{planckian}. Planckian conductivity \cite{patel} above
$T_{C}$ reinforces surviving CO "smectic" order characterized by
one-dimensional quantum transport subject to scatterings of mesoscopic physics
(transmission and reflections). This is related to other Planckian quantum
transport phenomena such as chiral interactions in the topological integer
quantum Hall effect (IQHE) \cite{physca}.

The Planckian phonon scattering rates are linear in temperature, approximated
using the thermal barrier, $k_{B}T$, as,
\begin{equation}
\frac{1}{\tau_{phonon}}=\alpha\frac{k_{B}T}{\hbar}, \label{linear1}%
\end{equation}
where $\alpha$ is on the order of unity \cite{patel}, multiplied by the number
of parallel stripes in the $CuO_{2}$ plane. The impurity scattering rates
formula (well-known Landauer conductivity \cite{physreports}), is governed by
the transmission and reflection of potential scatterers in 1D channels. Thus,
we can easily deduce \cite{physreports} the scattering rates as,
\begin{equation}
\frac{1}{\tau_{Coulomb}}=C\frac{e^{2}}{\hbar} \label{linear2}%
\end{equation}
where $C\simeq\varsigma\frac{1}{\eta}$ is inversely proportional to the mean
free path, $\eta$, between impurity scatterings, and $\zeta$\ is the number of
parallel stripes. Therefore, the total scattering rate is the sum,%
\begin{equation}
\frac{1}{\tau}=\frac{1}{\tau_{phonon}}+\frac{1}{\tau_{Coulomb}}
\label{linear3}%
\end{equation}
which is linear with respect to temperature.

This 1D formula is based on the approximation of the conducting stripes in the
2D $CuO_{2}$ plane as 1D parallel channels across the 2D CuO plane at
temperatures above the SC dome in the overdoped regions. Indeed, some
experimental works claim that CO survives at temperatures above the SC dome in
the overdoped regions. This symmetry-breaking CO survives above the $T_{c}$
although the holes are no longer superconducting leading to a strange metal
phase above the over-doped region of the SC dome. Thus, conducting stripes are
still present although the current-carrying holes are no longer
superconducting but are subject to phonon and impurity scatterings of 1D
mesoscopic physics. Therefore, the resistance formula of mesoscopic physics,
also known as the Planckian regime, is applied.

Experiments on antiferromagnetic--chain-mediated entanglement precisely
support this new entanglement mechanism and confinement. We expect that above
the superconducting dome, the strange metal behavior dominates because of the
persistence of parallel 1D conduction channels where non-superconducting
carriers at both ends of the antiferromagnetic-chain link reside. All of the
above strongly support the new pairing mechanism of RECHP, as a new and very
intuitive pairing mechanism in high-$T_{C}$ cuprates.

\subsubsection{Fan-out of strange metal behavior above the overdoped region}

The dome signifies that disordered RECHP pair transform to the low-energy
"smectic" CO configurations, such as the one illustrated in Figs. \ref{fig4}
and \ref{fig3}. We attribute the fan-out behavior above $T_{C}$ as due to
surviving "smectic" CO as enhanced by the in-plane electric fields, thus
maintaining the mesoscopic parallel 1D conducting channels above $T_{C}$,
though the hole conducting channel at both ends of the antiferromagnetic chain
link are no longer in phase or no longer move in unison with charge
$2\left\vert e\right\vert $ \cite{huang,patel, planckian}. However, we expect
the strange metal phase to be at temperatures above the overdoped region, that
is, starting above the dome peak and through the over-doped region of the SC
dome, where $\Delta_{C}=\Delta^{\ast}$ or $T_{C}=T^{\ast}$ as shown in Fig.
\ref{phaseD}, where the "smectic" or CO still survives as indicated in Fig.
\ref{schematic}.

\subsection{Beyond SC dome and Fermi liquid state}

In the overdoped region beyond optimal doping in the dome, $\Delta_{T^{\ast}%
}=\Delta_{C}$. Moreover, $T^{\ast}=T_{C}$ \ is coincident in the overdoped
region of the dome (in contrast to $T^{\ast}$ curve tangent to $T_{C}$ dome)
before the metallic phase, as shown in Fig. \ref{phaseD}. In the over-doped
regions, the weakened coupling caused by the shorter effective
antiferromagnetic-link (weaker confinement or EEF) between entangled holes,
statistically brought about by the increasing population of holes, will on
average start to dominate so that superconductivity starts to set at
temperatures lower than the optimal point. This decrease in $T_{C}$ which
coincides with $T^{\ast}$ will continue with further increase in doping
levels, until the \textit{measure} of entanglement approaches zero leading to
a Fermi liquid state, and the system eventually behaves as a conventional
paramagnetic conductor with a higher disorder or entropy.

\subsection{Electron-doped cuprate of the phase diagram}

Electron-doped cuprate superconductors are hole-driven superconductors owing
to the generation of holes with electron doping at oxygen sites \cite{e as
h,egenerateh}. These holes form Zhang-Rice singlets \cite{e as h} with
unpaired electrons at the Cu sites. Except for the holes located at the oxygen
sites rather than at the Cu sites, the formation of singlets, via the
Zhang-Rice mechanism, with the nearest member of the antiferromagnetic chain,
resembles a rehash of the RVB pairing mechanism. There appears to be nothing
to prevent them from pairing via the RECHP mechanism. This RECHP mechanism
results in a stronger coupling than the Zhang-Rice mechanism. Clearly, the
phase diagram for electron-doped cuprates exactly mirror those of hole-doped.
The start of the SC dome near the peak indicates a large number of generated
holes occurs once the electron doping attain a certain doping level. The phase
diagram shows a pseudogap for electron-doping levels before the peak of SC
dome, including the singularity of the $T^{\ast}$-linear curve at the SC dome,
as depicted in Fig. \ref{schematic} in the analysis for hole-doped. This
brings in the sought-after features of universality in the entire phase
diagram of high-$T_{C}$ cuprates.

The strength of the coupling $J^{\prime}$ between the hole at the oxygen site
and the electron spin at the Cu site is probably weaker, and the effective
antiferromagnetic chain link is shorter, resulting\ in a smaller $\Delta_{C}$
and hence $T_{C}$ than for hole-doped cuprates. Although, the BCS-like
treatment of itinerant electrons \cite{itinerant} may also contribute, their
low transition temperatures were not observed in the experiments and are not
shown in the phase diagram, except at the low $T_{C}$ tails of the phase
diagram of electron-doped cuprates.

The lack of a full pseudogap phase in Fig. \ref{phaseD}\ of electron-doped
cuprates is an indication of electron-doping-induced generation of entangled
holes, less entanglement measure or EEF, and/or higher resonating frequency
leading to a lower $T_{C}$ compared to hole-doped cuprates. The confinement
mechanism is crucial for explaning the linear pseudogap region of underdoped
cuprates as well as the $T^{\ast}$-singularity.

It is worth pointing out that the super-exchange energy, $J_{ij}<0$, is
approximately one to two orders of magnitude larger than the energy gap of the
BCS superconductor, that is, $10^{-3}$ to $^{10-2}$ eV compared to $10^{-4}$
eV for the energy gap of the BCS superconductor. Because $\Delta\left(
L_{eff}\left(  d\right)  \right)  $ for a given length is a summation of
superexchange $J_{ij}$, this could be several orders of magnitude stronger
than the BCS and RVB pairing.

\section{Concluding remarks}

The concept of entanglement in strongly correlated systems was hinted at in
\cite{anderson,ref9,ref11,kitaev}. However, the concept of confinement has
never been taken into consideration. The main point of this study is that the
SB "smectic" CO current-carrying pattern depicted in Figs.\ref{fig3} and
\ref{fig4} can readily explain the \textit{stripy} pattern of hole
superconductivity in high-$T_{C}$ cuprates. The rivers of superconductive
charge, spin-polarized and spin-unpolarized stripes, a version of spin-charge
separation, are natural consequences of our model, as well as the presence of
a spin gap in Figs.\ref{schematic} and \ref{phaseD}. The idea of confinement
also helps to elucidate the decrease in $T_{C}$ with overdoping, the decrease
in $T^{\ast}$ in the pseudogap phase, as well as the $T^{\ast}$-singularity at
the SC dome peak. The SB "smectic" CO which persists at $T>T_{C}$ predicts a
linear-$T$ resistivity of 1D mesoscopic physics of stripes in contrast with
conventional metals. For $T>T_{C}$, the holes at both ends of the
antiferromagnetic link no longer move in unison with charge $2\left\vert
e\right\vert $ but now move in an uncoordinated manner as independent 1D
parallel channels for hole mesoscopic transport as described by Eqs.
(\ref{linear1}) - (\ref{linear3}). In the following section, we cite
experiments that further support the foundation and predictions of our new
pairing-mechanism model.

\subsection{Analysis of spin texture experiments}

The following analysis relies on the CO and CE=0 phase depicted by Fig.
\ref{fig3} and Fig. \ref{fig4}. The spin texture dependence on the doping
level clearly signifies the dominant role of the \textit{long-range}
entanglement of hole dopants, as CO of independent Bell basis states, Fig.
\ref{fig3}, or a series of mixed long chains of triplet-singlet entangled
pairs, Fig. \ref{fig4}, depending on the dopant level and doping material in
the antiferromagnetic environment, as we shall see in what follows.

\subsubsection{Doping dependence of spin texture in high-$T_{c}$ cuprates}

Spin-resolved ARPES spectra of the spin texture of $Bi_{2}Sr_{2}CuO_{6+x}$
(Bi2212) and Pb-doped, $Bi_{2-x}Pb_{x})Sr_{2}CaCu_{2}O_{8+x}$ (Pb-Bi2212) were
obtained by Gotlieb, et al \cite{gotlieb}, Iwasawa, et al \cite{iwasawa} and
Lou, et al \cite{lou}. Iwasawa, et. al. has raised some of the difficulties in
SR-ARPES experiments and emphasized that because of the complexity of the spin
texture reported by Gotlieb, et al \cite{gotlieb}, the origin of spin
polarization in high-$T_{c}$ cuprates remains unclear. Iwasawa, et al.
\cite{iwasawa} differed from Gotlieb, et al. \cite{gotlieb}. Here we sense
some reproducibility issues due to the complex \textit{dynamical} origin of
the spin texture caused by the doping-dependent presence of spin-polarized and
spin-unpolarized conducting channels. This is discussed in connection with the
higher-order entanglement process shown in Fig. \ref{fig4}.

\subsubsection{Single-Layer $Bi_{2}Sr_{2}CuO_{6+x}$}

Lou, et al. \cite{lou} made some interesting observations. Two main trends are
observed in their data: (1) a decrease in the spin polarization from overdoped
to underdoped samples for both coherent and incoherent quasiparticles, and (2)
a shift of spin polarization from positive to negative as a function of
momentum. Positive and negative polarizations can occur as shown in Fig.
\ref{fig4}. The actual spin-polarization measurements may depend on the
geometry, orientation, or size of the sample.

The present consensus is that local structural fluctuations drive spin texture
in high-$T_{C}$ cuprate superconductors. In line with the local
crystal-symmetry breaking view proposed in Ref. \cite{lou}, we associate the
"smectic" configuration of entanglement, schematically shown in Fig.
\ref{fig4}, where triplet entanglements \textit{acting as emergent qubits}
\cite{perspective,buotbook2} are entangled as singlets and vice versa. \ These
higher-order processes result in a spin-polarized 1D channel. This type of
entanglement is likely dominant in the overdoped region, as suggested by
experiments \cite{lou}.

\subsubsection{\label{suppress}Suppression of spin polarization in Pb-doped
(Bi$_{2-x}$Pb$_{x}$)Sr$_{2}$CaCu$_{2}$O$_{8+x}$}

A striking reduction in spin polarization is observed in the coherent part of
the spectra for the Pb-doped sample with respect to Bi2212, with the imbalance
of the spin-up and spin-down intensities completely diminished \cite{currier}.
With either the absence or reduced local crystal symmetry breaking for
Pb-doped (Bi$_{2-x}$Pb$_{x}$)Sr$_{2}$CaCu$_{2}$O$_{8+x}$\cite{lou}, we
conclude that the configurational order of this boson system of degenerate
states defines a "smectic" pattern schematically depicted in Fig. \ref{fig3}.
In other words, the observed striking reduction in spin polarization is due to
the transition of the conducting channel dominance from the spin-polarized
channel shown in Fig. \ref{fig4} to the unpolarized configuration shown in
Fig. \ref{fig3}, as observed by Currier \cite{currier}.

Local structural fluctuations drive spin texture in high-temperature cuprate
superconductors. In line with the local crystal symmetry breaking view
proposed in Ref. \cite{lou}, we associate the higher-order process in the form
of entangled triplets, by acting as emergent qubits, which are now entangled
into a singlet pairing and vice versa. This is schematically illustrated in
Fig. \ref{fig4}. The resulting spin dynamics are the origin of the
doping-dependent complex spin texture found in the SR-ARPES experiments for
the overdoped region. However, in the underdoped region the decrease in the
polarization maybe due to the "smectic" CO consisting of a series of
independent or noninteracting alternate $\Phi$ and $\Psi$ entangled hole pairs
where the rivers of charge are unpolarized, Fig. \ref{fig3}.

A striking reduction of the spin polarization or spin texture for the Pb-doped
sample, with respect to Bi2212 \cite{currier} is due to onset of "smectic" CO
arrangement of alternate $\Phi$ and $\Psi$ entangled hole pairs where the
rivers of charge are not polarized, Fig. \ref{fig3} . Therefore, the source of
the spin texture is the local structural fluctuations. This is less for the
Pb-doped sample compared to that of Bi2212 where structural fluctuations
induce higher-order pairing, as depicted in Fig. \ref{fig4}.

As we have seen, this new pairing mechanism qualitatively explains the phase
diagram of both electron and hole-doped cuprates, notably the pseudogap, spin
gap, superconducting stripes, and strange-metal linear-$T$ behavior above the
overdoped regions of SC dome of hole-doped cuprates. Fanout of the strange
metal behavior at $T>T_{C}$ is attributed to the influence of the in-plane
electric fields in maintaining "smectic" CO of antiferromagnetic links with
doping concentration.

\begin{acknowledgments}
We thank Danilo Yanga, Gibson Maglasang and Allan R. Elnar for helpful
comments and for providing some of the references.
\end{acknowledgments}

\end{document}